# REPRODUCTIVE OUTCOME IN FEMALE WISTAR RATS TREATED WITH NHEXANE, DICHLOROMETHANE AND AQUEOUS ETHANOL EXTRACTS OF *Cucurbita pepo* SEED

**Anyanwu, C. F.[a*], Georgewill, O. A.[a] and Obinna, Victoria C.[b]**

a Department of Pharmacology, Faculty of Basic Clinical Sciences, University of Port Harcourt, P.M.B. 5323, Nigeria

b Department of Animal and Environmental Biology, Faculty of Science, University of Port Harcourt, Rivers State, Nigeria

**Corresponding Author**

Dr. Chinwe F. Anyanwu
Department of Pharmacology
Faculty of Basic Clinical Sciences
University of Port Harcourt
ORCID: 0000-0002-0632-6306
Email: Chinwe.anyanwu@uniport.edu.ng

## ABSTRACT

In developing countries, healthcare challenges and expensive infertility treatments has resulted in resurgent interest in medicinal plants. This study was designed to determine if *Curcubita pepo* seed can enhance female fertility, by assessing the reproductive outcome in female wistar rats treated with n-hexane (nHE), dichloromethane (DCM) and aqueous ethanol (Aq. Eth) extracts of *Curcubita pepo* seed. Total of 48 rats randomly grouped into 12 (n=4), were treated for 21 days by oral gavage as follows: A (control) = 0.5ml 20% tween 80 (vehicle); B (positive control) = 10mg/kg clomiphene citrate, C, D & E = 142.86, 285.71 and 428.57 mg/kg nHE; F, G & H = 142.86, 285.71 and 428.57 mg/kg DCM ; and I, J & K =142.86, 285.71 and 428.57 mg/kg Aq.Eth extracts. Group L (positive control 2) = 10mg/kg clomiphene citrate for 8 days. Following treatment, the rats were paired with males for mating, designating the confirmation day as gestational day 0 (GD 0). On GD 20, the animals were laparatomised and reproductive outcome was determined by assessing foetal weight, foetal crown-rump length, litter size, number of implantation and resorption sites. Results showed all extracts had no significant ($p > 0.05$) effect on the reproductive outcome indices. Clomiphene citrate significantly decreased reproductive outcome indices. In conclusion, *Cucurbita pepo* seed did not enhance the reproductive outcome of treated female rats at the doses and duration used in this study. This finding may serve as a springboard for future studies exploring the effect of *C.pepo* at different doses or durations.

## INTRODUCTION

Infertility is a very complex health challenge with physiologic, psychologic and economic dimensions (1). The simplest definition for infertility is the inability to conceive after one year of unprotected and uninterrupted sexual intercourse (2,3) or in the absence of any identified reproductive pathology (4). Globally, about 15 percent of couples suffer from infertility (5) and one in six couples suffer from infertility related issues during their reproductive age (Boivin *et al.,* 2009; Polis *et al.*, 2017; Nasimi *et al*., 2018).

Based on the findings of different studies, approximately 20–50 % of infertility is male caused, while 40% is associated with females and 25% is idiopathic (2,5,7,8). Female infertility has different aetiologies, which include ovarian diseases, cervical problems, congenital anomalies, endometriosis, dysfunction of the hypothalamus-pituitary-ovarian axis, tubal disorders, uterine pathologies, and systemic diseases (2,9).

In developing countries such as Nigeria, due to the inadequacies in healthcare, and the high cost of managing infertility, many women resort to the use of herbal therapies (8,10–12) which is usually affordable and readily available. This has therefore led to a resurgent interest in medicinal plants or herbal preparations that exhibit fertility properties.

*Curcubita pepo* L. [Pumpkin], a traditional plant commonly consumed in almost all parts of Eastern Nigeria, has been found to be among the common herbal remedies associated with reproductive health care (13,14). Pharmacologically, it has been reported to have the following activities such as anti-inflammatory, anti-hypercholesterolemia, anti-hypertensive, anti- parasitic, anti-carcinogenic, anti-oxidant, anti-bacterial, (15). Research findings on *C. pepo* seed shows that it contains different categories of phyto-constituents such as linoleic, Palmitic, stearic and oleic acids (16), flavonoids and alkaloids (17) which may be implicated in its pertinent medicinal properties.

Research findings have reported that pumpkin seed extract may recover the side effects of certain treatments on male reproductive health, possibly through preventing oxidative stress (18). The context provided by the findings on the protective effects of pumpkin seed extract in males adds further relevance to our investigation, prompting an exploration of whether similar protective effects or other reproductive outcomes can be observed in females. This study seeks to address the existing research gap and provide insights into the potential benefits and risks associated with the use of *Cucurbita pepo* in the area of female fertility and fecundity.

Historically, *Cucurbita pepo* (Pumpkin) has been reported to be used for diverse health purposes, including relief from nausea and vomiting of pregnancy (NVP), motion sickness, and as a nutritional supplement. The existing literature suggests that *Cucurbita pepo* is generally considered safe in pregnancy, as indicated by previous studies (13,14,19). However, despite this reported safety in pregnancy and the promising protective effects of *C. pepo* in males, the specific impact of *Cucurbita pepo* on female fertility remains an area that warrants further exploration. Therefore, the aim of this study was to evaluate the reproductive outcome in female Wistar rats treated with extracts (n-hexane, dichloromethane, and aqueous ethanol) of *Cucurbita pepo* seed, as an indicator for fertility.

## MATERIALS AND METHOD

### Plant Material and Authentication

Fresh fruits of *Cucurbita pepo* (pumpkin) were bought from Choba central market, Port Harcourt, Rivers state. It was packaged in a polythene bag and transported to the Department of Pharmacology, university of Port Harcourt. Authentication of the plant species was done by comparing it with the voucher specimen (Ref No. UPH/PSB/2021/071) available at the University Herbarium.

### Preparation of the extract

The procured fresh fruits of *Curcubita pepo* were cut open to harvest the shelled seeds which were dried under a shade at room temperature for four weeks and then deshelled. The deshelled seeds of *Cucurbita pepo* were weighed and ground to fine powder. The extraction process was done as illustrated by Harborne, (1998). Successive extraction method by cold maceration in solvent was done for 72 hours using 1.5 litres each of the three solvents - N-hexane, Dichloromethane and 70% aqueous ethanol. In each case, every 24 hours, there was fresh replacement of solvent. Extraction solvents were used in ascending order of their polarity, i.e. N-hexane, Dichloromethane and aqueous ethanol.

A five hundred grams (500g) of *Cucurbita pepo* seed powder was macerated in 1.5 Litres of n-hexane for 24hours. They were stirred at the expiration of the 24 hours and filtered first, with a Muslim fabric and the resultant solution was further filtered with Whatman's No. 1 filter paper. The Marc (residue) was soaked again with the same volume of n-hexane for another 24hours after which filtration of the Marc with a Muslim fabric is carried out, and thereafter, filtration with Whatman's filter paper was done. The resultant Marc was macerated again in another 1.5Litres of same solvent for 24hours again (i.e. a total of 72 hours of maceration) and the filtration process repeated. The filtrates were added together and concentrated with rotary evaporator (Model No; RE-52A) at $45^0$C *in vacuo* and then transferred to an evaporating dish and dried over a water bath (Digital thermostatic water bath, Jinotech instruments).

The resulting marc was dried to a constant weight for subsequent extraction with the second solvent - dichloromethane and then the third solvents - 70% aqueous ethanol in a similar manner to obtain the Dichloromethane *C.pepo* seed extract and 70% aqueous ethanol *C.pepo* seed extract respectively.

**Animals**

Adult female Wistar rats procured from the animal house of the Department of Pharmacology, University of Port Harcourt, Nigeria, were used for the study. Prior to the study, the experimental animals were acclimatized for two weeks. Clean drinking water and commercial feed from Top Feeds Nigeria were given to them *ad libitum*.

**Acute oral toxicity study**

This aspect of the study was done with a total of 54 female rats, in order to determine the $LD_{50}$ of the three extracts. The acute oral toxicity study for N-hexane, dichloromethane and aqueous ethanol extracts of *Cucurbita pepo* seed was performed using the method of Lorke, (1983). For each extract, 18 animals were used. For the N-hexane extract, in the first phase, 3 groups of 3 animals each, were given the extract at doses of 10mg/kg, 100mg/kg and 1000mg/kg by oral gavage, and observed for 24 hours. In the second phase, 3 groups of 3 animals each were given extract at doses of 1600mg/kg, 2900mg/kg and 5000mg/kg by oral gavage and observed for 24hours for death and other indicators of toxicity. Same procedure was repeated with the dichloromethane extract and the aqueous ethanol extract, each using 18 animals

**Experimental Design**

The doses of *Cucurbita pepo* seed extract used was as recommended by the Committee on Herbal Medicinal Products (HMPC) of the European Medicines Agency, (2012).

Forty-eight (48) female rats showing regular oestrous cycle with an average weight of 190g and 24 sexually mature male rats weighing about 200g were used for this study. The male rats were used only for mating. The female Wistar rats were randomly divided into 12 groups of four animals each for the following treatment:

Group A (Control)         –      0.5ml of 20% Tween 80.
Group B (positive control) –     10mg/kg of Clomiphene Citrate
Group C                   -      142.86mg/kg of N-hexane Extract
Group D                   -      285.71mg/kg of N-hexane Extract
Group E                   -      428.57mg/kg of N-hexane Extract
Group F                   -      142.86mg/kg of Dichloromethane Extract
Group G                   -      285.71mg/kg of Dichloromethane Extract
Group H                   -      428.57mg/kg of Dichloromethane Extract

| Group I | - | 142.86mg/kg of Aqueous ethanol Extract |
| Group J | - | 285.71mg/kg of Aqueous ethanol Extract |
| Group K | - | 428.57mg/kg of Aqueous ethanol Extract |
| Group L (positive control 2) - | | 10mg/kg of Clomiphene Citrate |

All treatments were administered daily by gavage for 21days (using a gavage syringe) except group L which was administered for 8 days duration. The rats were weighed every three days and the doses adjusted accordingly.

**Reproductive Outcome**

On day 21, at the end of extract administration, two females were put together with a male overnight in a cage. The presence of sperms in vaginal aliquot and / or vaginal plug was an indication that mating was established. This was referred to as gestational day (GD) 0 (23,24). On gestational day 20 (GD 20), the rats were anaesthetized, dissected; the uteri were excised and incised at the greater curvature of the horns. The reproductive outcome was assessed by evaluating the gestation and foetal indices according to Kagbo & Obinna, (2018) using the following parameters:

1. The total uterine implants and resorptions.
2. The mean crown rump length of the pups.
3. The mean litter size
4. The mean pup weight at gestational day 20 (GD 20)

**Statistical Analyses**

Statistical analyses were done with SPSS 21; the data were represented as mean ± SEM, and assessed using one-way Analysis of Variance (ANOVA) and Tukey post-hoc test. The significance level was set at $p<0.05$.

**RESULTS**

**Acute Toxicity Study**

There was no mortality, morbidity or other apparent signs of toxicity shown with the acute toxicity study, at the doses used. This was a proof that all three extracts were not noxious at maximum dose of 5000mg/kg. However, the Committee on Herbal Medicinal Products (HMPC) of the European Medicine Agency, 2012, (22) recommended that *C. pepo* be administered to adult human (70kg) at a dosage range of 10 – 30g. Thus, this was adopted for the studies, which gave rise to 143mg/kg, 286mg/kg, 429mg/kg doses of n-hexane, dichloromethane and aqueous ethanol extracts of *Cucurbita pepo* seed as used in this study.

## Effect of n-Hexane Extract of *C.pepo* seed on Reproductive outcome of Female Wistar Rats

Table 1 shows that all the doses of n-hexane seed extract of *C. pepo* as used in this study caused no significant (P>0.05) variation in the reproductive outcome of treated female wistar rats relative to the normal control (group A) as assessed by foetal weight of the pups, foetal crown rump length of the pups, litter size of the rat, number of implantation as well as resorption sites on the rats' uterine horns.

Table 2 shows that the Dichloromethane extract of *Cucurbita pepo* seed had no significant (P>0.05) effect on the reproductive outcome (foetal weight, foetal crown rump length, litter size and number of implantation site as well as resorption sites) of rats treated for 21 consecutive days when compared with the control (group A).

Aqueous ethanol extract of *Cucurbita pepo* seed at the doses of 142.86mg/kg, 285.71mg/kg and 428.57mg/kg did not produce any significant (P>0.05) effect on the reproductive outcome (foetal weight, foetal crown rump length, litter size and number of implantation & resorption sites) of test rats in comparison with the control (Group A).

Similarly, the pups from the test groups were normal; no sign of pathology on the pups, placenta and umbilical cord was observed (figure 2).

**Table 1: Effect of n-hexane extract of *C. pepo* seed on the Reproductive Outcome of female wistar rats**

| Groups | Foetal Weight(g) | FCRL(cm) | Litter size (No) | Implantation sites (No) | Resorption sites (No) |
|---|---|---|---|---|---|
| A | 3.56±0.08[#] | 3.67±0.07[#] | 6.25±1.84[#] | 6.50±1.66[#] | 0.25±0.25 |
| B | 0.00±0.00 | 0.00±0.00 | 0.00±0.00 | 0.00±0.00 | 3.25±3.25 |
| C | 3.29±0.10[#] | 3.65±0.06[#] | 8.25±0.48[#] | 8.25±0.48[#] | 0.00±0.00 |
| D | 3.31±0.11[#] | 3.71±0.05[#] | 5.50±0.87[#] | 7.00±1.22[#] | 1.50±1.50 |
| E | 3.33±0.12[#] | 3.68±0.08[#] | 8.25±0.25[#] | 8.25±0.25[#] | 0.00±0.00 |

Results are given as Mean ± SEM fir each group. Statistical evaluation was done by one-way ANOVA, followed by Tukey's post-hoc test. Experimental groups are compared with group A (Normal Control) and group B (Positive Control – Clomiphene citrate). *p<0.05 was considered as significant versus the Normal control (Group A); [#]p < 0.05 was considered significant versus the positive control (Group B).

**Table 2: Effect of Dichloromethane extract of *C. pepo* seed on the Reproductive Outcome of female Wistar Rats.**

| Groups | Foetal | FCRL(cm) | Litter size | Implantation | Resorption |
|---|---|---|---|---|---|

|   | Weight(g) |   | (No) | sites (No) | sites (No) |
|---|---|---|---|---|---|
| A | 3.56±0.08# | 3.67±0.07# | 6.25±1.84# | 6.50±1.66# | 0.25±0.25 |
| B | 0.00±0.00 | 0.00±0.00 | 0.00±0.00 | 0.00±0.00 | 3.25±3.25 |
| F | 3.73±0.51# | 3.83±0.24# | 8.50±1.26# | 8.75±1.25# | 0.25±0.25 |
| G | 3.30±0.12# | 3.66±0.05# | 9.00±0.71# | 9.00±0.71# | 0.00±0.00 |
| H | 3.82±0.12# | 3.81±0.03# | 6.75±0.85# | 7.00±0.71# | 0.00±0.00 |

Results are given as Mean ± SEM for each group. Statistical evaluation was done by one-way ANOVA, followed by Tukey's post-hoc test. Experimental groups are compared with group A (Normal Control) and group B (Positive Control – Clomiphene citrate). *$p<0.05$ was considered as significant versus the Normal control (Group A); #$p < 0.05$ was considered significant versus the positive control (Group B).

**Table 3: Effect of Aqueous ethanol extracts of *Cucurbita pepo* seed on Reproductive outcome of Female Wistar Rats**

| Groups | Foetal Weight(g) | FCRL(cm) | Litter size (No) | Implantation sites (No) | Resorption sites (No) |
|---|---|---|---|---|---|
| A | 3.56±0.08# | 3.67±0.07# | 6.25±1.84# | 6.50±1.66# | 0.25±0.25 |
| B | 0.00±0.00 | 0.00±0.00 | 0.00±0.00 | 0.00±0.00 | 3.25±3.25 |
| I | 4.16±0.82# | 3.94±0.23# | 6.50±1.55# | 6.50±2.40# | 1.00±1.00 |
| J | 3.19±0.30# | 3.54±0.16# | 4.75±1.32 | 4.75±1.32 | 0.00±0.00 |
| K | 3.87±0.03# | 3.97±0.04# | 6.50±0.29# | 6.50±0.29# | 0.00±0.00 |

Results are given as Mean ± SEM fir each group. Statistical evaluation was done by one-way ANOVA, followed by Tukey's post-hoc test. Experimental groups are compared with group A (Normal Control) and group B (Positive Control – Clomiphene citrate). *$p<0.05$ was considered as significant versus the Normal control (Group A); #$p < 0.05$ was considered significant versus the positive control (Group B).

**Effect of Clomiphene citrate on Reproductive outcome of Female Wistar Rats**

From the result, out of the four (4) clomiphene citrate treated-rats (positive control – group B) administered for 21 days, no pup was seen at GD 20 on laparotomy, even though mating was confirmed by presence of vaginal plug or sperm cells in vaginal smear in all, instead 13 pinpoint areas of resorption (as seen in figure 1) were seen on the two (2) uterine walls of just one (1) out of the four (4) rats. Similarly, when clomiphene citrate was administered to four (4) female rats for 8 days (group L) and mating was confirmed, at gestation day 20 (GD 20) following laparotomy, only one (1) out of the four (4) rats had just a pup.

**Table 4: Effect of Clomiphene citrate administered for 21 days on GROUP B (Positive Control) rats**

| Animals | Foetal Weight(g) | FCRL(cm) | Litter size (No) | Implantation sites (No) | Resorption sites (No) | Remarks |
|---|---|---|---|---|---|---|
| 1 | - | - | - | - | - | No pup |
| 2 | - | - | - | - | - | No pup |
| 3 | - | - | - | - | - | No pup |
| 4 | - | - | - | - | 13 | No pup, 13 Pinpoint areas of resorption seen on the uterine horns |

Table 5: Effect of Clomiphene citrate administered for 8 days on GROUP L (Positive Control 2)

| Animals | Foetal Weight (g) | FCRL (cm) | Litter size (No) | Implantation sites (No) | Resorption sites (No) | Remarks |
|---|---|---|---|---|---|---|
| 1 | - | - | - | - | - | No pup |
| 2 | - | - | - | - | - | No pup |
| 3 | - | - | - | - | - | No pup |
| 4 | 3.71 | 3.8 | 1 | 1 | 0 | 1 pup |

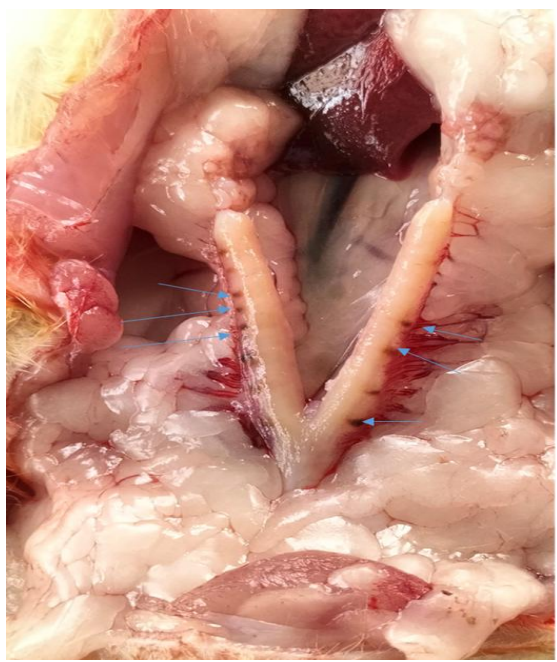
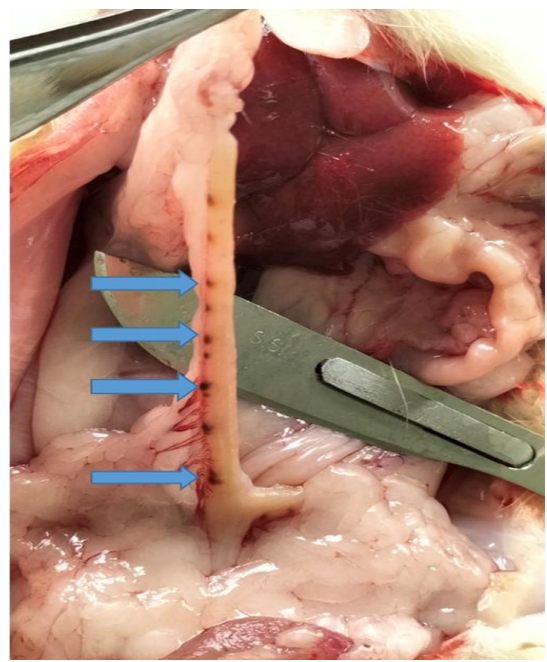

Figure 1: Picture of the uterine horns of clomiphene-treated rat (21 days duration) showing pinpoint resorption sites (arrows)

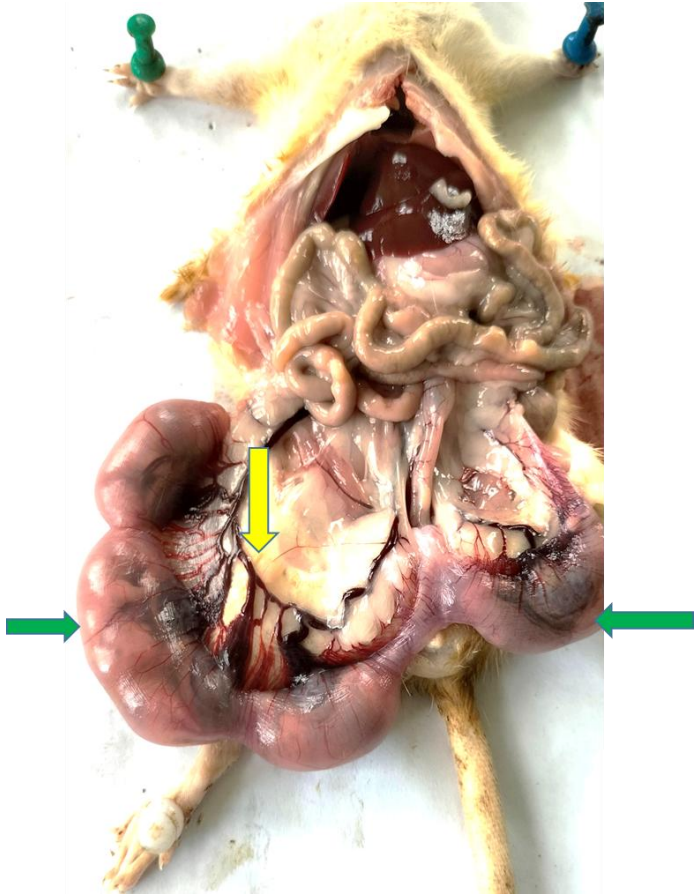

Figure 2: The picture of a laparatomised pregnant rat showing the gravid left and right uterine horns (green arrows), and the engorged blood vessels (yellow arrow) supplying the uterine horns and the placenta

**DISCUSSION**

The study investigated the reproductive outcomes of female Wistar rats treated with three extracts *of Cucurbita pepo* seed, alongside clomiphene citrate as a positive control. The results of this study indicated that the three extracts of *Cucurbita pepo* seed did not have any significant effect on the indices for assessing the reproductive outcome of treated rats. In contrast, when compared to other studies, particularly those focusing on the protective effects of pumpkin seed extract on male reproductive health post-chemotherapy (CP) treatment, a different narrative emerged (18). The pumpkin seed extract study demonstrated potential in alleviating the adverse effects of chemotherapy, specifically in terms of epididymal histology and sperm parameters. The demonstrated ability to counteract oxidative stress suggests a possible mechanism through which pumpkin seed extract may have contributed to preserving epididymal health and maintaining sperm quality in the face of CP-induced challenges (18). In the same line, studies by Ofoego *et al*., (2017) revealed remarkable improvements in testicular parameters following the administration of pumpkin seed extract in *Azadirachta Indica* (Neem) induced reproductive toxicity. The differential responses between male and female rats may suggest intricate interactions between *Cucurbita pepo* seed extracts and the complex hormonal and physiological pathways governing reproductive health. However, the protective effects observed in the pumpkin seed extract study underscore the multifaceted nature of herbal remedies, revealing potential benefits in specific reproductive contexts.

Results from this study shows that Clomiphene citrate, a selective estrogen receptor modulator (SERM) used as the first line of treatment in managing infertility in normally oestrogenized, anovulatory women (WHO group II), did not yield a positive reproductive outcome. The findings of this study aligns with the concerns raised by Homburg, (2005) regarding the effectiveness of Clomiphene citrate. It was reported that the effectiveness of Clomiphene citrate as measured by the single live birth rate of approximately 25% for starters, leaves room for improvement (27). The gaps between ovulation and pregnancy rates, as well as cases of non-response to Clomiphene citrate, are recognized issues that have prompted the exploration of alternative treatment options (28,29). However, our findings contrasted with that of Ogbuehi *et al.,* (2015), which stated that clomiphene citrate led to a significant increase in the number of pups littered by treated rats, highlighting the variability in treatment responses across different experimental contexts.

Successful initiation of pregnancy involves various factors including ovulation of a mature oocyte, production of competent sperm, proximity of sperm and oocyte in the reproductive tract, fertilization of the oocyte, transport of the conceptus into the uterus, and implantation of the embryo into a properly prepared, healthy endometrium. Dysfunction in any of these steps can lead to infertility (Ruder *et al.,* 2008). Despite the demonstrated successful initiation of pregnancy, there was no significant difference in reproductive outcomes between the treatment group and the normal control. The choice of the three different solvents (n-hexane, dichloromethane, and aqueous ethanol) for the extraction of *Cucurbita pepo* seed compounds is crucial. It has been established that different solvents have different extracting powers which eventually affects the extract yield, availability of bioactive compounds or phytochemical constituents as well as the pharmacological activities of the plant material (31,32). Each of these solvents extracts different types of compounds, and their varying polarities can impact which phytochemicals are obtained. According to Anyanwu *et al*., (2022), only Palmitic, stearic, and linoleic compounds and their derivatives were bioactive compounds found in all three extracts. Research findings by Kim *et al*., (2012) revealed the major fatty acids found in *C.pepo* seeds grown in Korea were Palmitic, stearic, oleic and linoleic acids. The presence of essential fatty acids such as, linoleic acid and α-linolenic acid and their derivatives which balances female reproductive hormones and also aids in lubricating the mucous membrane (34), was found in all three extracts (Aqueous ethanol, dichloromethane and n-hexane) of *Cucurbita pepo* seed.

Nonetheless, based on the details presented in tables 1-3, there were no discernible variations in reproductive outcome measures such as fetal weight, fetal crown-rump length (FCRL), litter size, and the number of implantation sites when compared to the normal control. The lack of significance in these reproductive outcome parameters may suggest potential limitations associated with dosage and treatment duration. As the first of its kind, this study therefore, lays the groundwork for further research in exploring the detailed mechanisms underlying the protective effects of pumpkin seed extract.

The pinpoint areas of resorption observed in clomiphene citrate-treated rats for a duration of 21 days (as seen in figure 1), may be linked to a decrease in the expression of CD98 on the endometrial epithelium. CD98, a type II glycoprotein, is crucial for amino acid and hormonal transport and is expressed in ovarian, placental, and testicular tissues. It plays a significant role in the receptivity of the endometrium during implantation. Decreased expression of CD98, as seen in the clomiphene citrate group, may result in decreased hormonal and amino acid transportation in the endometrial tissue, potentially contributing to implantation failure (35,36).

**Conclusion**
In conclusion, the study indicates that *Cucurbita pepo* seed extracts did not enhance the fertility of female wistar rats as shown in the non-significant effect on the reproductive outcome indices in treated rats. Clomiphene citrate, despite its widespread use in infertility treatment, did not yield positive results. Based on our findings, we recommend conducting future studies that specifically investigate the role of CD98 in implantation failure, aiming to elucidate detailed pathways and molecular interactions. Additionally, refining treatment protocols, considering factors such as higher dosage and prolonged duration of exposure, is crucial for further studies on the reproductive outcomes of *C.pepo* seed.

**COMPETING INTERESTS**
Authors have declared that no competing interests exist.

**ACKNOWLEDGEMENTS**
The authors are grateful to Lively stone diagnostic Laboratory, for providing the facility for the hormonal assay.